\documentclass[useAMS,usenatbib,usegraphicx]{mn2e}
\usepackage{latexsym,graphicx,natbib,amssymb}

\usepackage{color}

\newcommand\kpc{{\rm\,kpc}} 
 
\newcommand\msun{\rm\,M_\odot}

\newcommand{\aj}{AJ}
\newcommand{\apj}{ApJ}
\newcommand{\apjl}{ApJ}
\newcommand{\apjs}{ApJS}
\newcommand{\aap}{A\&A}

\newcommand{\mnras}{MNRAS}
\newcommand{\nat}{Nature}

\newcommand{\pasp}{PASP}

\newcommand{\aplett}{ApL}

\title[Mass segregation in Palomar\,14]
 {Mass segregation in the outer halo globular cluster Palomar~14}
      
\author[M. J. Frank et al.]
{{\Large Matthias J. Frank$^{1,2}$\thanks{E-mail: mfrank@lsw.uni-heidelberg.de}, Eva K. Grebel$^{1}$ and Andreas H.
W. K\"upper$^{3,4}$}\\
$^1$ Astronomisches Rechen-Institut, Zentrum f\"ur Astronomie der Universit\"at
Heidelberg, M\"onchhofstrasse 12~-~14, D-69120 Heidelberg, Germany \\
$^2$ Landessternwarte, Zentrum f\"ur Astronomie der Universit\"at
Heidelberg, K\"onigsstuhl 12, D-69117 Heidelberg, Germany \\
$^3$ Department of Astronomy, Columbia University, 550 West 120th Street, New York, NY 10027, USA\\
$^4$ Hubble Fellow
}

\date{Accepted ????. Received ?????; in original form ?????}

\pagerange{\pageref{firstpage}--\pageref{lastpage}} \pubyear{2011}

\begin{document}   

\maketitle

\label{firstpage}

\begin{abstract}
We present evidence for mass segregation in the outer-halo globular cluster Palomar\,14, which is intuitively unexpected since its present-day two-body relaxation time significantly exceeds the Hubble time. Based on archival Hubble Space Telescope imaging, we analyze the radial dependence of the stellar mass function in the cluster's inner $39.2$\,pc in the mass range of  $0.53\leq m\leq 0.80\msun$, ranging from the  main-sequence turn-off  down to a $V$-band magnitude of 27.1\,mag. The mass function at different radii is well approximated by a power law and rises from a shallow slope of  $0.6\pm0.2$  in the cluster's  core  to a slope of  $1.6\pm0.3$ beyond $18.6$ \,pc. This is seemingly in conflict with the finding by \citet{2011ApJ...737L...3B}, who interpret the cluster's non-segregated population of (more massive) blue straggler stars, compared to (less massive) red giants and horizontal branch stars, as evidence that the cluster has not experienced dynamical segregation yet. We discuss  how both results can be reconciled.
Our findings indicate that the cluster was either primordially mass-segregated and/or used to be significantly more compact in the past. For the latter case, we propose tidal shocks as the mechanism driving the cluster's expansion, which would imply that Palomar\,14 is on a highly eccentric orbit. Conversely, if the cluster formed already extended and with primordial mass segregation, this could support an accretion origin of the cluster.
\end{abstract}


\section{Introduction}
\label{sec:intro}
The vast majority of the Galaxy's globular clusters (GCs) have apparent half-mass two-body relaxation times much shorter than their respective ages \citep[e.g.][2010 edition]{1996AJ....112.1487H}. In these clusters, at least in their centers where the density is highest, two-body relaxation has shaped the distribution of stars: in two-body encounters, massive stars tend to lose kinetic energy to lower-mass stars and, as a result, massive stars sink into the cluster's center, whereas low-mass stars gain energy allowing them to populate orbits further away from the cluster's center. This is observed as mass segregation, i.e. more massive stars show a more concentrated radial distribution than lower-mass stars. As a consequence, the mass function of the cluster becomes shallower in the cluster center but also globally, as stars at large radii are more easily lost to the Galactic potential \citep{1997MNRAS.289..898V} and the cluster's mass-to-light ratio decreases \citep[e.g.][]{2009A&A...500..785K}.

However, in a few GCs, such as the massive clusters NGC\,2419 and NGC\,5139 ($\omega$Cen) and several low-mass, but extended, outer halo clusters, the dynamical half-mass relaxation time $T_{\mathrm{relax},rh}$ is comparable to or even exceeds the Hubble time. The mass segregation time-scale for a star of mass $m$ is given $<\!m\!>/m\times T_\mathrm{relax}$ \citep{1969ApJ...158L.139S}, where the average stellar mass $<\!m\!>$ in GCs is $\sim\!0.2-0.3\msun$. Therefore dynamical effects like low-mass star depletion or mass segregation should not have affected the (luminous) main-sequence stars ($m\!\sim\!0.7\msun$) in these clusters much, assuming that the cluster parameters such as half-mass radii and cluster masses have not changed dramatically within the last few Gyr and hence that their present-day two-body relaxation times are a representative measure of the relaxation time-scale for their entire lifetime. 

One of these clusters is Palomar\,14 \citep[Pal\,14; also known as AvdB after its discoverers][]{1960PASP...72...48A}.  \citet{2011ApJ...737L...3B} recently reported that the distribution of blue straggler stars in Pal\,14 is not centrally concentrated with respect to less massive HB and RGB stars in this cluster. Blue stragglers (BS) are hydrogen-burning stars that are brighter and bluer than the main-sequence turn-off of their host population and are believed to form either through direct collisions \citep{1976ApL....17...87H} or through mass-transfer in binary systems \citep{1964MNRAS.128..147M}. The radial distributions of BS in massive globular clusters \citep[e.g.][]{2006MNRAS.373..361M,2007ApJ...663..267L} as well as the detection of two distinct sequences of BS in M\,30 \citep{2009Natur.462.1028F} suggests that both mechanisms are at work in dense GCs. In a diffuse cluster like Pal\,14, the low stellar density and therefore low probability of collisions makes it likely that BS are primarily due to mass-transfer in primordial binary systems. Regardless of the formation channel, the masses of BS, or at least the total masses of their progenitor binary systems, likely exceed those of any other single star that can be identified in a cluster's color-magnitude diagram \citep[e.g.][]{1997ApJ...489L..59S,2011Natur.478..356G}. Hence,  the distribution of BS with respect to lower-mass tracers is widely used as an indicator for mass segregation and the state of dynamical evolution of GCs \citep[e.g.][]{2012Natur.492..393F}. Accordingly,  \citet{2011ApJ...737L...3B} interpreted their finding as proof that two-body relaxation has not yet affected Pal\,14, not even in its center, and that the cluster is not mass-segregated, in agreement with the expectation from its long present-day half-mass relaxation time of $\sim20$\,Gyr \citep{2011ApJ...726...47S}. 

Pal\,14 's dynamical state has recently also  received considerable attention in the context of testing modified Newtonian dynamics  \citep[MOND;][]{1983ApJ...270..371M}. The cluster's  large Galactocentric distance of $\sim$66\,kpc \citep{2011ApJ...726...47S} in combination with its low mass and density, and therefore the low external and internal acceleration experienced by its stars, make it an excellent test case for this theory \citep{2005MNRAS.359L...1B,2010MNRAS.401..131S,2011A&A...527A..33H}.  In  this context, \citet{2009AJ....137.4586J} measured radial velocities of 16 cluster members and found a good agreement between the cluster's photometric and dynamical mass in classical Newtonian dynamics. Their measured  velocity dispersion is significantly lower than predicted in MOND \citep{2009MNRAS.395.1549H}, although MOND cannot be ruled out from these data \citep{2010A&A...509A..97G}. \citet{2010ApJ...716..776K} reanalyzed the \citet{2009AJ....137.4586J} radial  velocities  including a heuristic treatment of binaries and mass segregation, and argued that Pal\,14 has to have a very low binary fraction in order to be  compatible with the low observed velocity dispersion. In a Monte-Carlo analysis  of the data that included  the effects of binaries, of the external field and of velocity anisotropy, \citet{2012ApJ...744..196S} found that the cluster is compatible with Newtonian dynamics also when the constraint of the binary fraction is relaxed to $\la\!30$ per cent.  Tidal tails around the cluster, first detected at low significance in Sloan Digital Sky Survey \citep[SDSS;][]{2009ApJS..182..543A} data by \citet{2010A&A...522A..71J}  and  confirmed in deeper wide-field imaging by \citet{2011ApJ...726...47S} may further complicate the interpretation of the small sample of radial velocities as unbound stars may contaminate the sample \citep[e.g.][]{2011MNRAS.413..863K}.

Pal\,14 was also the subject of the first full collisional $N$-body simulations of a GC over its entire lifetime by \citet{2011MNRAS.411.1989Z}. These authors found that, while two-body relaxation does have an effect on the cluster's stellar mass function, the effect from two-body relaxation is too weak such that the shallow present-day mass function slope\footnote{Throughout this paper, the sign convention is such that positive $\alpha$ corresponds to negative power-law exponent, i.e. the \citet{1955ApJ...121..161S} IMF has a slope of $\alpha=+2.35$.} of $\alpha=1.3\pm0.4$ measured by \citet{2009AJ....137.4586J} cannot be reproduced with models starting with a \citet{2001MNRAS.322..231K} initial mass function (IMF). \citet{2011MNRAS.411.1989Z} demonstrated that the observations can, in principle, be reproduced by models with primordial mass segregation, but that the necessary degree of mass segregation has to be very high. Alternatively, they argued that the cluster must have lost a good fraction of its low mass stars in the early gas-expulsion phase, referring to an effect studied by \citet{2008MNRAS.386.2047M}. However, also this latter scenario would require that the cluster must have formed with a high degree of primordial mass segregation. Hence, in any case Pal\,14 should be mass-segregated nowadays, since otherwise the observational evidence for a flattened present-day mass function of this cluster is at odds with our current understanding of stellar dynamics or with the assumption of a universal IMF. 

In this contribution, we use archival Hubble Space Telescope (\textit{HST}) data reaching down to $\sim4$\,mag below the main-sequence turn-off to test whether the cluster is mass-segregated in order to shed light on its dynamical state. The paper is structured as follows: we present the observational data we use and describe our data analysis in Section~\ref{sec:obs}. In Section~\ref{sec:res} we report our results on the state of mass segregation in Pal\,14, followed by a discussion of our findings in Section~\ref{sec:disc}, and we end with a short summary of our work in Section~\ref{sec:sum}. 

\section{Observations and Literature Data}
\label{sec:obs}
Our analysis is based on archival \textit{HST} imaging, obtained with the Wide-Field Planetary Camera 2 (WFPC2) in program GO 6512 (PI: Hesser). The data were first published in the literature by \citet{2008AJ....136.1407D} and were also used to measure the slope of the cluster's overall central mass function  by \citet{2009AJ....137.4586J}, who did not attempt quantify mass segregation due to the relatively small WFPC2 field of view. The data consist of F555W ($V$) and F814W ($I$) band images with the cluster approximately centered on the WF3 chip of the camera. With exposure times of $4\times160$\,s, $1\times900$\,s and $7\times1000$\,s (amounting to 8540\,s in total) in F555W, and $4\times230$\,s, $4\times1100$\,s, $2\times1200$\,s, and $2\times1300$\,s (10320\,s in total) in F814W, this is the deepest available imaging of the cluster's center, reaching out to a radius of 1.9\,arcmin, or 39.2\,pc (assuming a distance of $71$\,kpc, see Section~\ref{sec:populationparameters}).

\begin{figure}
\includegraphics[width=84mm]{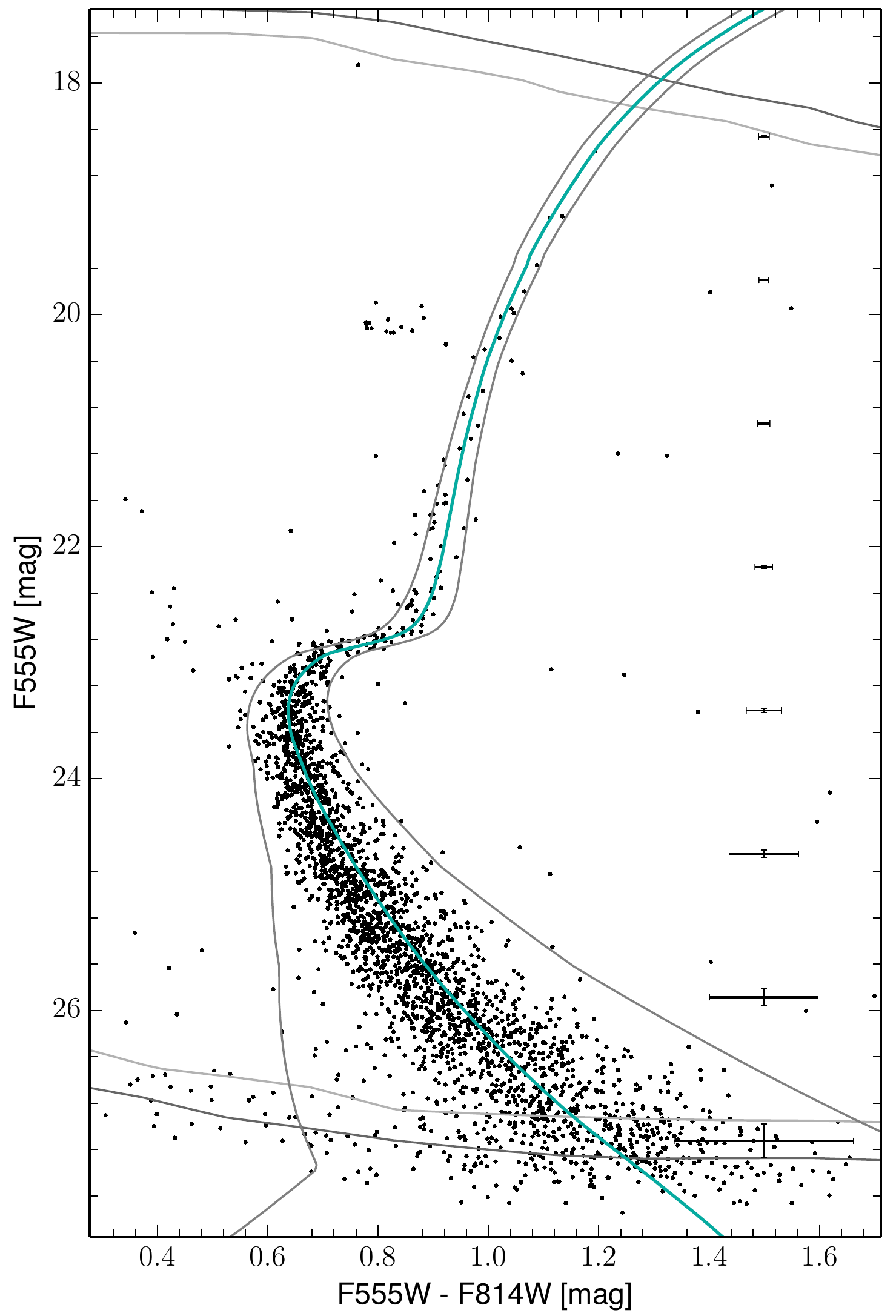}
\caption{Observed color-magnitude diagram of Pal\,14. Error bars on the right represent the photometric errors derived from artificial star tests. The gray lines at the faint and bright ends represent the 80\% (light gray) and 50\% (dark gray) completeness contours; at the bright end (F555W$\la$18\,mag), completeness declines due to saturation even in the shorter exposures. The isochrone, taken from the Dartmouth Stellar Evolution Database \citep{2008ApJS..178...89D} corresponds to an age of 11.5\,Gyr, [Fe/H]$=-1.5$\,dex and [$\alpha$/Fe]$=+0.2$\,dex (see Section~\ref{sec:populationparameters}). Gray curves to the left and to the right of the isochrone represent the color limits used for our analysis of the cluster's mass function.}
\label{fig:CMD}
\end{figure}

The data were reduced in the same way as described in more detail in section 2.3 of \citet{2012MNRAS.423.2917F}. PSF-fitting photometry was obtained using the \textsc{HSTPHOT} package \citep{2000PASP..112.1383D}. \textsc{HSTPHOT} was initially run on the individual images and the resulting catalogs were cross-matched in order to refine the image registration. The refined image registration was used for a cosmic ray rejection with \textsc{HSTPHOT}'s crmask task, and as an input for the photometry from all frames using one of the deep F555W frames as reference image.  \textsc{HSTPHOT} provides the photometry obtained from each frame, as well as a combined photometric catalog with average magnitudes for each filter, which we used.  The following quality cuts were applied to the resulting all-frames photometric catalog to select clean stellar detections (for details, see the \textsc{HSTPHOT} user manual): a type parameter of 1 (i.e. a stellar detection), abs(sharpness)$<0.2$, $\chi<2.0$, and in both filters a crowding parameter $<1.5{\rm\,mag}$ and a statistical uncertainty in the magnitude $<0.2{\rm\,mag}$. The resulting color-magnitude diagram, containing 3201 stars, is shown in Fig.~\ref{fig:CMD}.

 \begin{figure}
\includegraphics[width=84mm]{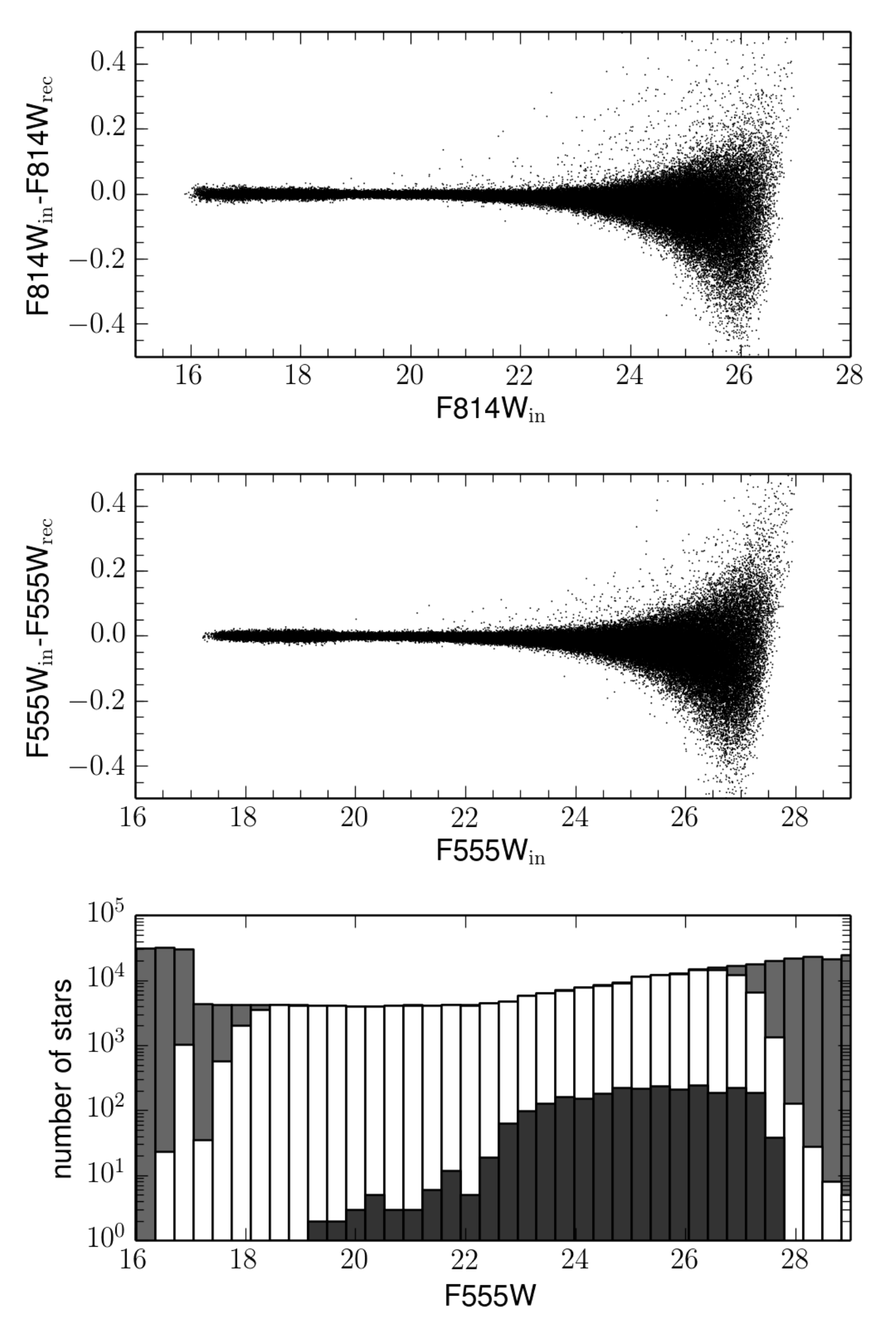}
\caption{The upper two panels show the photometric errors derived from the artificial star tests as a function of magnitude in the F814W (top panel) and F555W (middle panel) filter. Shown is the difference between inserted magnitude and recovered magnitude based on $\sim200000$ artificial stars that lie within the color selection shown in Fig.\ref{fig:CMD}.  The lower panel shows the luminosity function in the F555W filter of observed stars (dark gray histogram), and inserted (light gray) and recovered (white) artificial stars within the color selection, to illustrate that the artificial star tests sample the whole range of magnitudes very well.}
\label{fig:photerrors}
\end{figure}

\begin{figure}
\includegraphics[width=84mm]{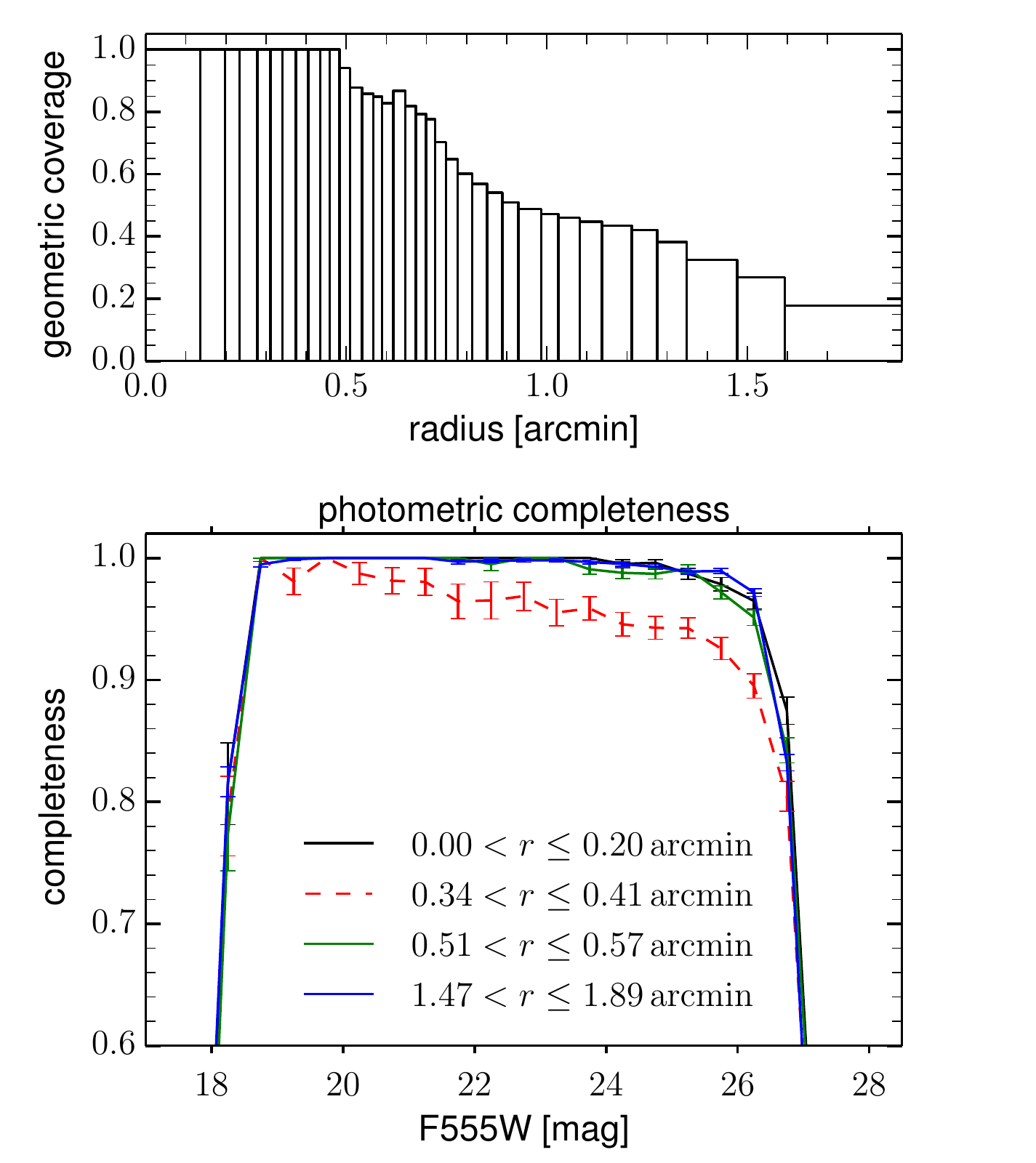}
\caption{Geometric coverage and photometric completeness of the WFPC2 data as a function of radius from the cluster center. The upper panel shows the geometric coverage of our photometric catalog, i.e. the area covered by the WFPC2 pointing in a given radial annulus divided by the total area of that annulus. The annuli were defined to contain each one 36th of the observed stars in our catalog. The lower panel shows the photometric completeness obtained from artificial star tests as a function of F555W magnitude (evaluated in bins of 0.5\,mag) in  four representative radial ranges. Completeness only marginally varies with radius, indicating that crowding does not strongly affect the photometry in this sparse cluster.  There is a  dip in completeness at $r\!\sim\!0.35$\,arcmin  (red dashed curve) that  is caused by a very bright foreground star  located  at this distance from the cluster center.}
\label{fig:completeness}
\end{figure}

\textsc{HSTPHOT} was also used to perform artificial star tests with  $\sim1.4\times10^6$  fake stars in order to estimate the photometric uncertainties and completeness. Artificial stars were distributed  homogeneously on the  WFPC2  chips (\textsc{HSTPHOT} option 4096) and in the range $16\,\le\,V\,\le\,29$\,mag and $-1.5\,\le(V-I)\,\le\,4.5$\,mag in color-magnitude space. \textsc{HSTPHOT} transforms this into equivalent \textit{HST}-system $F555W$ and $F814W$ magnitudes and, while covering the entire grid, generates more artificial stars in regions of the CMD with a high density of observed stars (cf. the \textsc{HSTPHOT} User's Guide). Since for our subsequent analysis we selected only artificial stars falling close to the isochrone (see below), the exact distribution of inserted artificial stars is unimportant and this mechanism simply yields a more efficient sampling of the relevant color-magnitude regions. The lower panel of Fig.~\ref{fig:photerrors}, which shows the luminosity functions of observed stars, as well as inserted and recovered fake stars, illustrates that the whole range of observed magnitudes is covered very well by the artificial star tests. \textsc{HSTPHOT} inserts artificial stars into each image using the previously constructed PSF and attempts to re-find the inserted stars and measure their brightness.  At any given time, \textsc{HSTPHOT} adds only one single artificial star to all frames and then tests whether this star can be recovered. This procedure avoids self-crowding within the artificial star tests. In \textsc{HSTPHOT} artificial stars are considered detected, if they are found within 3 pixels (4.5 pixels on the PC chip) of their inserted position and if no other star within that radius dominates the flux at this position \citep[see][for a detailed description of the \textsc{HSTPHOT} artificial star procedure]{2006ApJS..166..534H}.  We applied the  same quality cuts as used for the observed stars  to the resulting  artificial star catalog.  Photometric uncertainties in a given region of the CMD and on the sky were then estimated from the differences between inserted and recovered magnitudes. Error bars on the right-hand side of Fig.~\ref{fig:CMD} represent the uncertainty in magnitude and color as a function of magnitude for main-sequence and RGB stars. For this and our analysis of the mass-function we used only the subset of $\sim200000$ artificial stars that lie within  the color selection indicated by the gray lines to both sides of the isochrone in the same figure. These color limits were chosen to exclude stars that deviate in color from the locus of the adopted isochrone (see Section~\ref{sec:populationparameters}) by more than 3$\sigma_\mathrm{col}$, where $\sigma_\mathrm{col}$ is the color uncertainty derived from the artificial star results in the corresponding region of the CMD.  The upper two panels of Fig.~\ref{fig:photerrors} show the distribution of photometric errors (i.e. inserted-recovered magnitude) as a function of magnitude in both filters derived from artificial stars that fall within the color selection around the isochrone.

The photometric completeness was estimated from the ratio of the number of recovered to the number of inserted artificial stars. The completeness, within the color selection, as a function of F555W magnitude, at four different distances  from the cluster center,  is shown in the lower panel of Fig.~\ref{fig:completeness}. It is apparent that crowding does not or only marginally affect the photometry, as the completeness is essentially independent of the distance from the cluster center, apart from a dip in completeness at $\sim0.35$\,arcmin  that is  caused by a bright foreground star in the field.

The geometric coverage of the WFPC2 photometry, shown in the upper panel of Fig.~\ref{fig:completeness}, was quantified as the ratio of the area of covered pixels in a given radial annulus around the cluster's center to the total area of that annulus. The covered pixels were selected on a distortion-corrected frame combined with \textsc{multidrizzle} \citep{2006HSTc.conf..423K} by the requirement that the covered pixels receive at least 25 per cent of the total exposure time in both filters and be included in the reference frame used for photometry \citep[see][]{2012MNRAS.423.2917F}. 
The stellar positions in the photometric and artificial star catalogs were transformed to the coordinate system of the distortion-corrected, drizzled frame and stars falling on pixels marked as `not covered' were removed from both catalogs to avoid border effects. For the cluster's center, we used the coordinates determined by \citet{2006A&A...448..171H}, which agree within the uncertainties with the center determined by \citet{2011ApJ...737L...3B}. Since the absolute astrometric accuracy of the \textit{HST} is limited by the accuracy of its guide star catalog, we registered the distortion-corrected coordinate frame to the SDSS DR7 catalog \citep{2009ApJS..182..543A}, before transforming the coordinates of the center to pixel coordinates in our photometric catalog.

\subsection{Foreground contamination}
\label{sec:foreground}
\begin{figure}
\includegraphics[width=84mm]{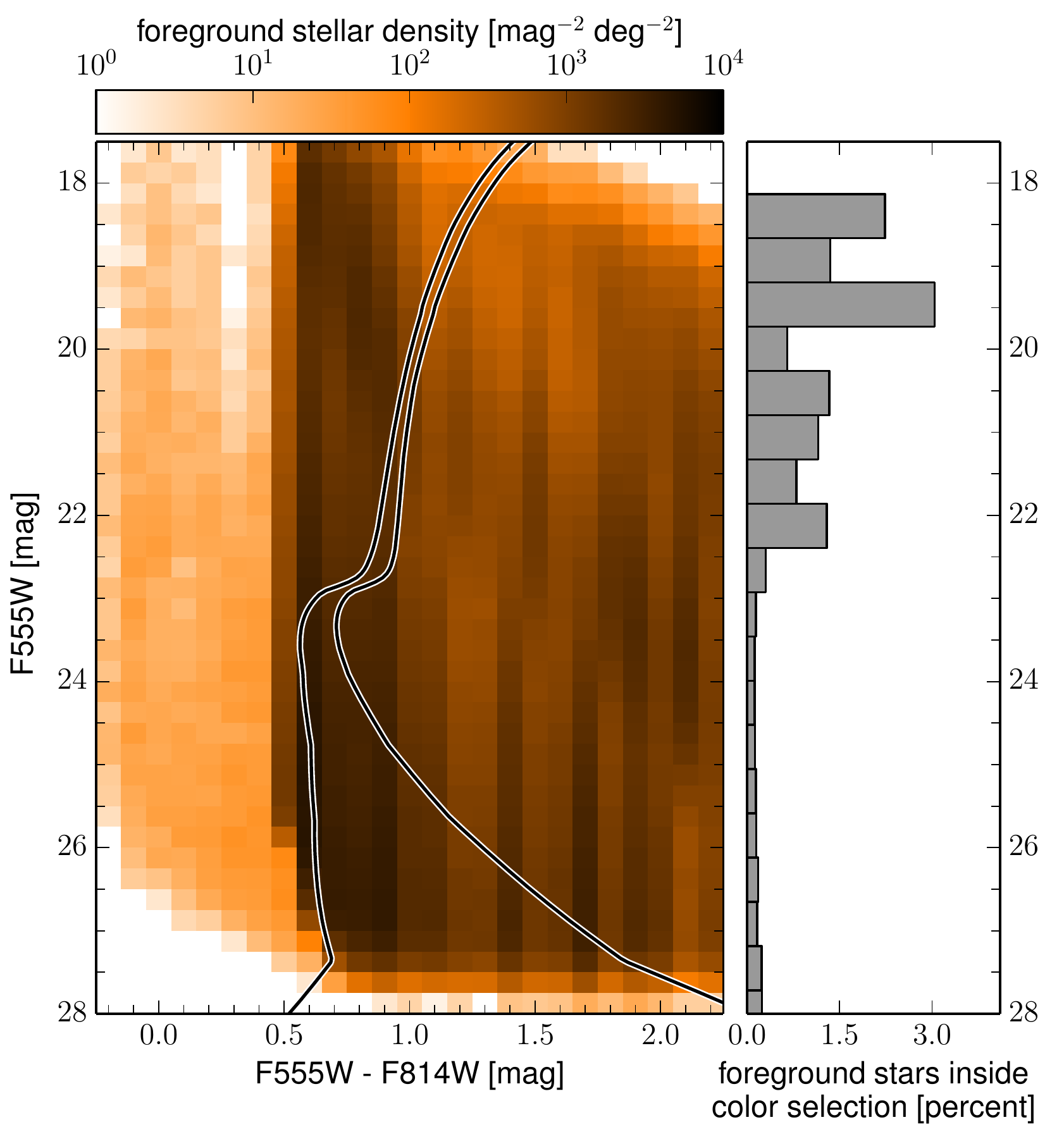}
\caption{The left panel shows the density of foreground stars in the color-magnitude diagram. The photometric uncertainties and completeness inferred from artificial star tests were imposed on the Besan\c{c}on foreground catalog. The density of foreground stars was evaluated in bins of $0.25$\,mag in F555W and $0.1$\,mag in color. The color scale indicates the density in units of foreground stars per square magnitude in the CMD and square degree on the sky. Black-on-white lines correspond to the region of the CMD used to estimate the mass function of Pal\,14. Within these color limits, the fraction of expected contaminants in the photometric catalog, averaged over $0.5$\,mag in F555W is shown in the right-hand panel. The fraction is well below 1 per cent on the main sequence and up to 3 per cent on the sparsely populated RGB.}
\label{fig:foreground}
\end{figure}

Since the WFPC2 photometry covers only the inner region of the cluster and no nearby comparison fields are available, we used the Besan\c{c}on model of the Galaxy \citep{2003A&A...409..523R} to estimate the expected contamination by foreground stars in our field. We queried the model for stars out to 200$\kpc$ in the direction of Pal\,14. For better number statistics, we used a solid angle of 50 square degrees and the model's `small field' mode that simulates all stars in the same line of sight, so that any spatial variation in the foreground that could be present in such a large field is neglected. The remaining model parameters, such as the extinction law and spectral type coverage, were kept at their default values. 

We transformed the $V$ and $I$ magnitudes of the resulting synthetic foreground catalog to F555W and F814W magnitudes by inverting the \citet{1995PASP..107.1065H} WFPC2 filter to $U\!BV\!RI$ transformations. The photometric uncertainties and completeness derived from artificial star tests were imposed on the foreground catalog using the procedure described in more detail in \citet{2012MNRAS.423.2917F}. The radial variation of these properties was taken into account by assuming the foreground stars to be distributed homogeneously over the sky and performing the procedure on radial sub-samples of artificial and foreground stars. 

The left panel of Fig.~\ref{fig:foreground} shows the density of foreground stars in our pointing, evaluated in bins of $0.25$\,mag in F555W and $0.1$\,mag in color and scaled to units of stars per mag in color and magnitude in the CMD and per square degree on the sky. Given the small effective area of the WFPC2 pointing of 4.72 square arcmin the number of expected foreground stars in our catalog is low. The ratio of expected foreground stars to observed stars inside our color selection, averaged over bins of 0.5\,mag in F555W, is below $\la\!3$ percent over the whole range of luminosities and even lower on the main sequence, as shown in the right-hand panel of Fig.~\ref{fig:foreground}.  Therefore, we chose to not correct the main-sequence star counts for foreground stars.

\subsection{Adopted stellar population parameters}
\label{sec:populationparameters}
In Section~\ref{sec:massfunction} we will use an isochrone as a relation between stellar magnitude and mass, and thus require the knowledge of the distance, reddening, age and chemical composition of the cluster. There are two photometric studies of the cluster's population parameters based on the same WFPC2 data we use here. \citet{2008AJ....136.1407D} analyzed the photometry transformed to the Johnson-Cousins $V$ and $I$ filters, and applied offsets in the $V$ and $I$ zeropoints to match the ground-based standard star photometry of \citet{2000PASP..112..925S}. In this system, they obtained a best-fitting Dartmouth isochrone \citep{2008ApJS..178...89D} for a distance of $79$\,kpc with [Fe/H]=$-1.5$\,dex, [$\alpha$/Fe]=$+0.2$\,dex and an age of 10.5\,Gyr. \citet{2009AJ....137.4586J} used the WFPC2 instrumental magnitudes and the same library of Dartmouth isochrones and adopted the spectroscopic metallicity of [Fe/H]$=-1.5$\,dex from \citet{1996AJ....112.1487H}, which in turn is based on measurements by \citet{1985ApJ...293..424Z} and \citet{1992AJ....104..164A}. They found a best-fitting isochrone with [$\alpha$/Fe]=$+0.2$\,dex, an age of 11.5$\pm0.5$\,Gyr, and a distance of $71\pm1.3$\,kpc.

From ground-based photometry, \citet{2006A&A...448..171H} derived an age of 10\,Gyr, [Fe/H]=$-1.5$\,dex, [$\alpha$/Fe]$=+0.3$\,dex and a distance of $77$\,kpc based on Yonsei-Yale isochrones \citep{2002ApJS..143..499K}. \citet{2011ApJ...726...47S} adopted [Fe/H]$=-1.6$\,dex and [$\alpha$/Fe]=$+0.3$\,dex and found a distance of $71\pm2$\,kpc as well as an older age of $13.2\pm0.3$~Gyr based on Padova isochrones \citep{2008A&A...482..883M}. While this age seems at odds with differential age dating, which suggests that Pal\,14 is several Gyr younger than classical old GCs \citep{1997AJ....113..682S,2008AJ....136.1407D}, differences in the absolute age may also stem from differences in the `age scale' of different sets of isochrones \citep[see e.g.][]{2008AJ....135.1106G,2009ApJ...694.1498M}. 

Finally, from high-resolution spectroscopy of red giants, \citet{2012A&A...537A..83C} recently derived a metallicity of [Fe/H]=$-1.3$\,dex (the mean of their [Fe\textsc{I}/H] and [Fe\textsc{II}/H] values) and [$\alpha$/Fe]=$+0.3\pm0.2$\,dex.

For our analysis in the WFPC2 instrumental magnitude system, we adopt the parameters derived by \citet{2009AJ....137.4586J} in this system and specifically also their reddening of $E(\mathrm{F555W}-\mathrm{F814W})=0.063$\,mag as well as their derived apparent distance modulus of $(m-M)_{\mathrm{F555W}}=19.45$\,mag. The adopted values are well within the scatter of parameters found by other authors, but we keep in mind that there is some scatter, even when the parameters are derived from the same data and with the same set of isochrones. We will discuss the influence of these uncertainties on the mass function in Section~ \ref{sec:massfunctionsystematics}.

\section{Results}
\label{sec:res}

\subsection{Luminosity function in the center and outer region}
\label{sec:lumfunc}
\begin{figure}
\includegraphics[width=84mm]{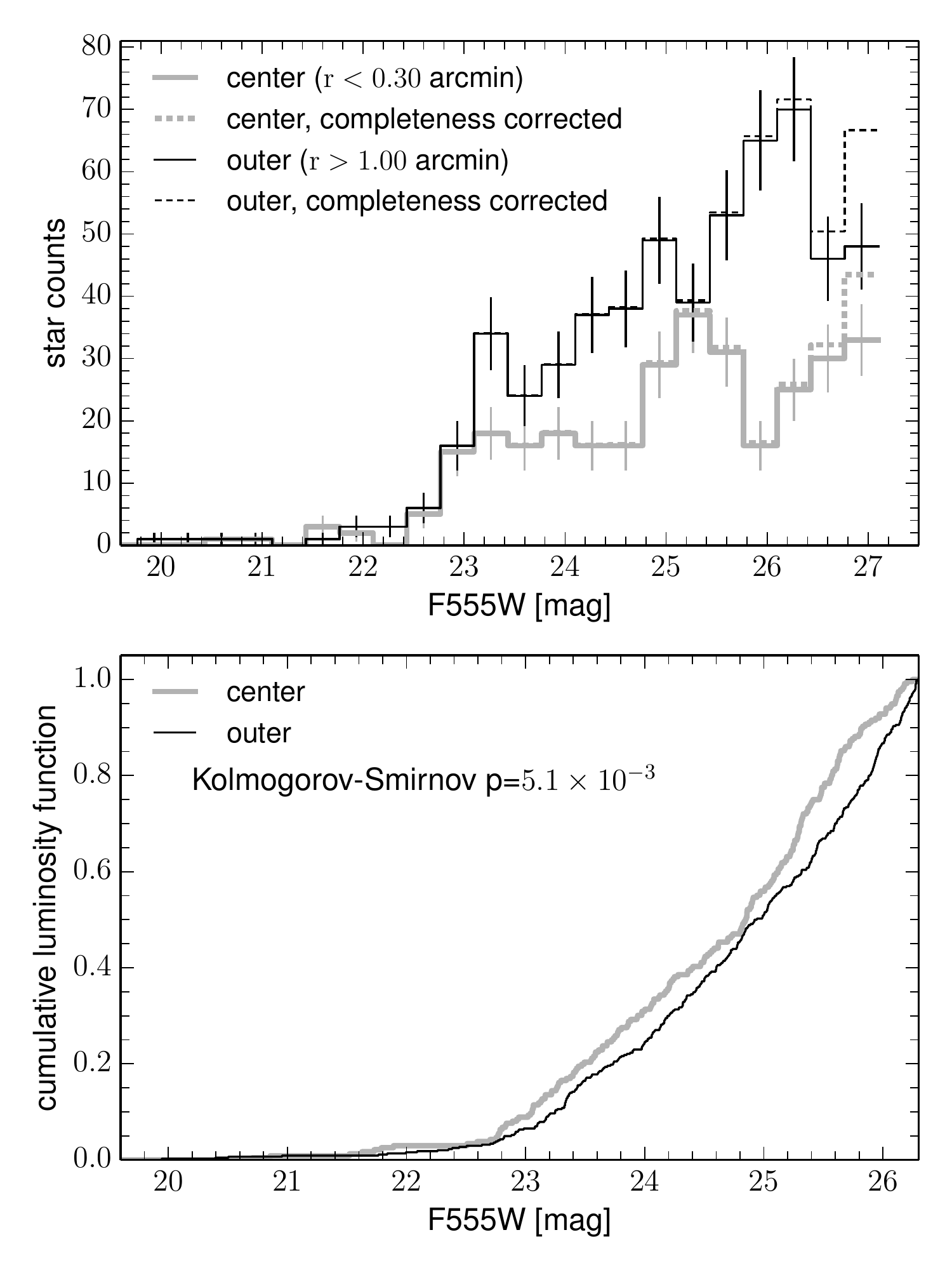}
\caption{The upper panel shows the observed luminosity functions in bins of 1/3\,mag in F555W for stars inside the color selection shown in Fig.~\ref{fig:lumfunc} in the center (thick gray lines; $r<0.3$\,arcmin) and the outer region covered by our catalog (thin black lines; $1<r<1.9$\,arcmin). The dashed gray and black lines correspond to the luminosity functions corrected for photometric incompleteness and expected foreground stars. It can be seen that while both samples contain similar numbers of stars around $\sim\!23$\,mag, just above the main-sequence turn-off, the outer sample contains more faint stars than the central region. The lower panel shows the cumulative luminosity distribution functions of observed stars brighter than 26.3\,mag in the central and outer region, where both samples are approximately complete. A Kolmogorov-Smirnov test yields a probability of 0.5\,per cent that the two samples are drawn from the same luminosity distribution. }
\label{fig:lumfunc}
\end{figure}

As a first diagnostic for the presence of mass segregation, we show in the  upper panel of Fig.~\ref{fig:lumfunc} the luminosity functions of stars falling into the color selection shown in Fig.~\ref{fig:CMD} in the innermost and outermost regions covered by our catalog. The central region, represented by the gray thick histogram, was defined as $r<0.3$\,arcmin, in order to avoid the region around the bright foreground star at $r\sim0.35$\,arcmin (see the lower panel of Fig.~\ref{fig:completeness}). The outer region (thin black line) was defined as $r>1$\,arcmin, where, as we will see in the next Section, the mass function appears to start to steepen. It is apparent that both the central and the outer sample contain a similar number of stars around 23\,mag in F555W, just above the main-sequence turn-off ($23.4{\rm\,mag}$). At fainter magnitudes, however, the luminosity function in the outer region appears to be more populated than in the central region, indicating the presence of mass segregation. The upper panel of Fig.~\ref{fig:lumfunc} also shows the luminosity functions corrected for photometric incompleteness as thick gray and thin black dashed lines for the central and outer regions, respectively. The completeness corrections were estimated from the ratio of recovered to inserted artificial stars inside the color selection shown in Fig.~\ref{fig:CMD} in the inner and outer region. They thus represent a crude overall completeness correction in the respective radial ranges; in the analysis in the following Section, a more precise, \emph{local} correction for photometric incompleteness and geometric coverage will be applied. We also calculated the luminosity function of foreground stars expected according to the Besan\c{c}on model from Section~\ref{sec:foreground} within the effective areas covered by the central and outer samples, of 0.3\,arcmin$^2$ and 2.34\,arcmin$^2$, respectively. As already expected from the right-hand panel of Fig.~\ref{fig:foreground}, the number of predicted foreground contaminants is very low. Their F555W luminosity function in the outer (inner) region rises almost monotonically from $0.006$ ($0.0008$) stars per $0.33$\,mag bin at $18.6$\,mag, to 0.2 (0.02) stars per bin at 26.9\,mag. Compared to the number of observed stars the contamination is therefore negligible and was not subtracted from the observed luminosity function.

In order quantify the apparent difference of the two luminosity functions shown in the upper panel of Fig.~\ref{fig:lumfunc}, we used a two-sided, two-sample Kolmogorov-Smirnov test to compare the cumulative (unbinned and not completeness-corrected) luminosity functions in both samples. For this, we used only stars brighter than 26.3\,mag, since these are largely unaffected by photometric incompleteness. The cumulative distribution functions are shown in the lower panel of Fig.~\ref{fig:lumfunc}. The Kolmogorov-Smirnov test yields a probability of only 0.5\,per cent that the observed luminosity functions of the central and outer region stem from the same underlying distribution. 

\subsection{Mass segregation}
\label{sec:massfunction}
We determined the stellar mass function in the cluster in the mass range  $0.53\leq m\le 0.80\msun$, corresponding to stars  fainter than the main-sequence turn-off with $23.4{\rm\,mag}\la{\rm F555W}\la27.1{\rm\,mag}$, inside the color selection  that is  shown in Fig.~\ref{fig:CMD} and defined in Section~\ref{sec:obs}. Photometric completeness, which decreases at the faint end, is still $\ga70$ per cent in the lowest mass bin at all radii. The color selection removed likely foreground stars  and blue stragglers. To each of the selected stars, we assigned a mass based on the adopted isochrone by interpolating the magnitude-mass relation defined by the isochrone to the star's measured F555W magnitude.

While the photometric completeness only marginally varies with distance from the cluster center, both the geometric coverage and the stellar density do vary as a function of radius (Fig.~\ref{fig:completeness}). Since we are interested in the mass function within a given radial range in the cluster, rather than in our catalog, it is necessary to apply a radially varying correction to the number of observed stars to ensure a correct weighting when combining stars from different radii to calculate the mass function. Correcting for the missing area in a given radial annulus necessarily has to assume that the cluster's stellar distribution is axially symmetric around the center. This is justified, given that a distortion of the isophotes that marks the onset of the tidal tails is visible only at radii $\ga5$\,arcmin \citep{2011ApJ...726...47S,2011ApJ...737L...3B}, while our catalog extends out to only $1.9$\,arcmin. We divided our photometric catalog into $n$ radial bins around the cluster center, and for each bin calculated the corrections for missing area and for photometric completeness as a function of F555W magnitude (see Fig.~\ref{fig:completeness}). The bins were chosen such that each of them contained one $n$th of the observed stars, optimal in terms of the Poissonian uncertainties of the star counts. $n$ has to be large enough to capture the variation of the geometric coverage and stellar density with radius, but should not be chosen larger than necessary to avoid statistical fluctuations caused by low numbers of artificial stars in the individual radial bins. We chose $n=36$ bins for our analysis after verifying that increasing the number of subdivisions further did not influence any more the derived overall mass function or the integrated stellar mass within $1.9$\,arcmin.

With only  2120  observed stars within the color selection and the magnitude limits reported above, the mass function in the individual 36 radial bins (containing  $\sim$59  stars each) is poorly constrained. We therefore re-combined several of these bins after the correction for geometric coverage and photometric completeness and measured the mass function. 

\begin{figure}
\includegraphics[width=84mm]{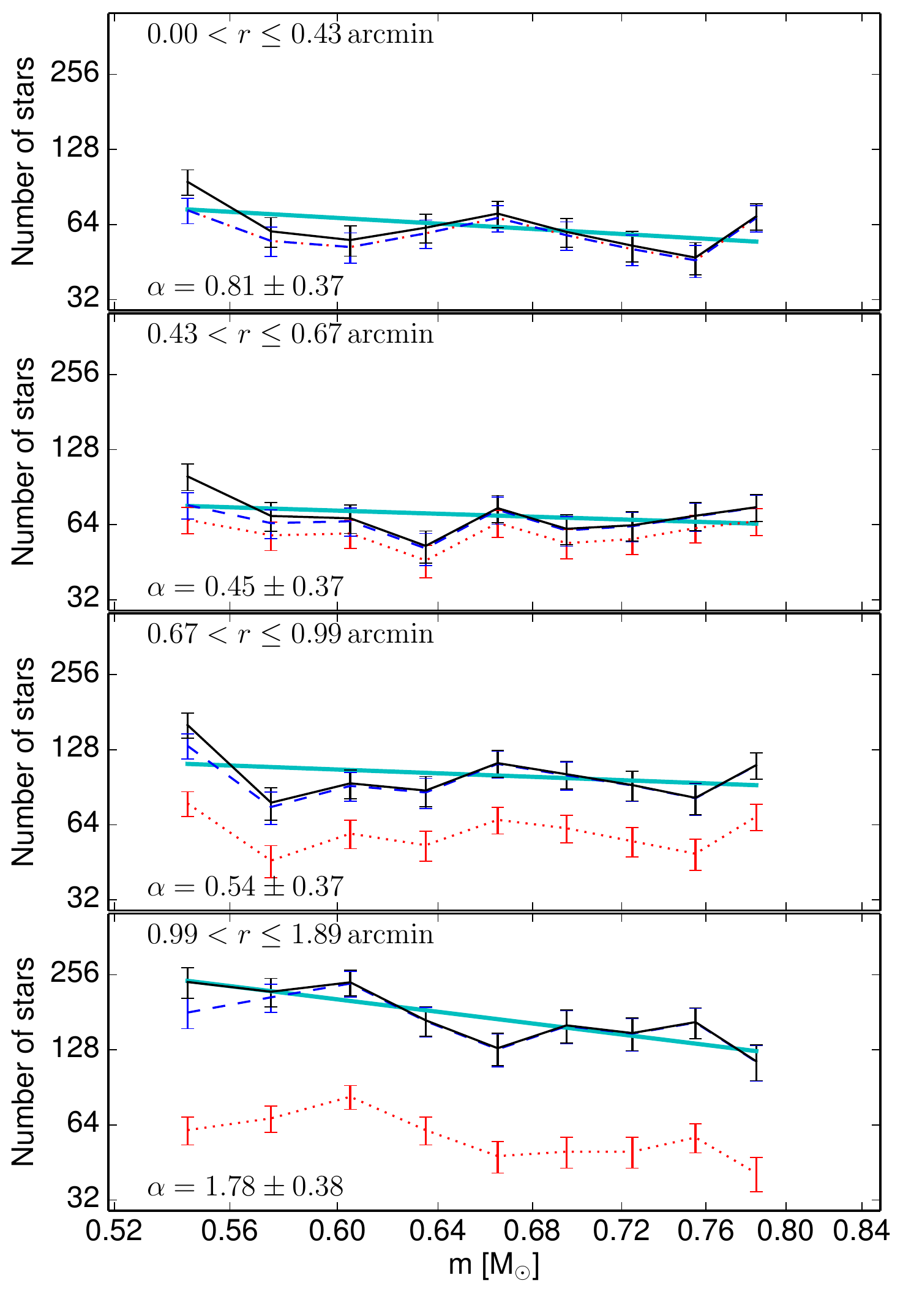}
\caption{The mass function in radial bins containing one fourth of the observed stars each; the radial ranges are reported at the top of each panel. Red dotted lines correspond to the raw star counts, blue dashed lines to the star counts after correction for geometric coverage and black lines to the star counts additionally corrected for photometric incompleteness. The best-fitting power-law mass function, obtained with the maximum-likelihood scheme described in the text is shown as cyan line. The best-fitting mass function slope $\alpha$ and its $1\,\sigma$ uncertainty is reported at the bottom of each panel. In all radial ranges, the mass function is well reproduced by a power-law and its slope is significantly shallower than a \citet{2001MNRAS.322..231K} mass function in this range of masses ($\alpha=2.3$).}
\label{fig:masssegregation}
\end{figure}

\begin{table}
\caption[]{Observed and completeness corrected mass function in four radial ranges as shown in Fig.~\ref{fig:masssegregation}. Listed are the mass bin center M, the number of observed stars N$_\mathrm{obs}$, the geometric coverage correction $f_\mathrm{geo}$, the photometric completeness $f_\mathrm{phot}$, and the number of stars corrected for geometric and photometric completeness N$_\mathrm{calc}$.}
\begin{tabular}{|r|r|r|r|r|r|}
\hline \hline
M [$\msun$] & N$_\mathrm{obs}$ & $f_\mathrm{geo}$ & $f_\mathrm{phot}$& N$_\mathrm{calc}$\\
\hline
\multicolumn{6}{c}{$0.00 <r\le 0.43\,\mathrm{arcmin}$}\\
0.54 & 73 & 1.00 & 0.77 & 94.9\\
0.57 & 55 & 1.00 & 0.92 & 60.1\\
0.60 & 52 & 1.00 & 0.94 & 55.5\\
0.63 & 59 & 1.00 & 0.95 & 62.2\\
0.66 & 68 & 1.00 & 0.96 & 70.8\\
0.69 & 58 & 1.00 & 0.97 & 59.8\\
0.72 & 51 & 1.00 & 0.97 & 52.7\\
0.75 & 46 & 1.00 & 0.98 & 47.2\\
0.78 & 68 & 1.00 & 0.99 & 69.0\\
\hline
\multicolumn{6}{c}{$0.43 <r\le 0.67\,\mathrm{arcmin}$}\\
0.54 & 67 & 0.87 & 0.77 & 99.9\\
0.57 & 58 & 0.89 & 0.94 & 69.4\\
0.60 & 59 & 0.89 & 0.97 & 67.9\\
0.63 & 46 & 0.89 & 0.98 & 52.6\\
0.66 & 65 & 0.89 & 0.99 & 74.5\\
0.69 & 54 & 0.89 & 0.99 & 61.6\\
0.72 & 56 & 0.89 & 0.99 & 63.6\\
0.75 & 62 & 0.90 & 0.99 & 69.5\\
0.78 & 66 & 0.88 & 1.00 & 75.2\\
\hline
\multicolumn{6}{c}{$0.67 <r\le 0.99\,\mathrm{arcmin}$}\\
0.54 & 78 & 0.59 & 0.83 & 160.7\\
0.57 & 46 & 0.61 & 0.96 & 78.5\\
0.60 & 59 & 0.64 & 0.98 & 93.8\\
0.63 & 53 & 0.61 & 0.98 & 87.8\\
0.66 & 67 & 0.60 & 0.99 & 113.1\\
0.69 & 62 & 0.61 & 0.99 & 102.0\\
0.72 & 55 & 0.60 & 1.00 & 92.3\\
0.75 & 49 & 0.60 & 1.00 & 82.1\\
0.78 & 69 & 0.62 & 1.00 & 111.0\\
\hline
\multicolumn{6}{c}{$0.99 <r\le 1.89\,\mathrm{arcmin}$}\\
0.54 & 61 & 0.34 & 0.75 & 239.3\\
0.57 & 68 & 0.33 & 0.95 & 218.8\\
0.60 & 83 & 0.35 & 0.99 & 238.6\\
0.63 & 61 & 0.37 & 0.99 & 168.0\\
0.66 & 48 & 0.37 & 0.99 & 129.9\\
0.69 & 50 & 0.31 & 0.99 & 160.4\\
0.72 & 50 & 0.34 & 1.00 & 149.4\\
0.75 & 57 & 0.35 & 1.00 & 165.3\\
0.78 & 41 & 0.36 & 1.00 & 115.1\\
\hline \hline
\end{tabular}
\label{tab:massfuncs}
\end{table}

\begin{figure}
\includegraphics[width=84mm]{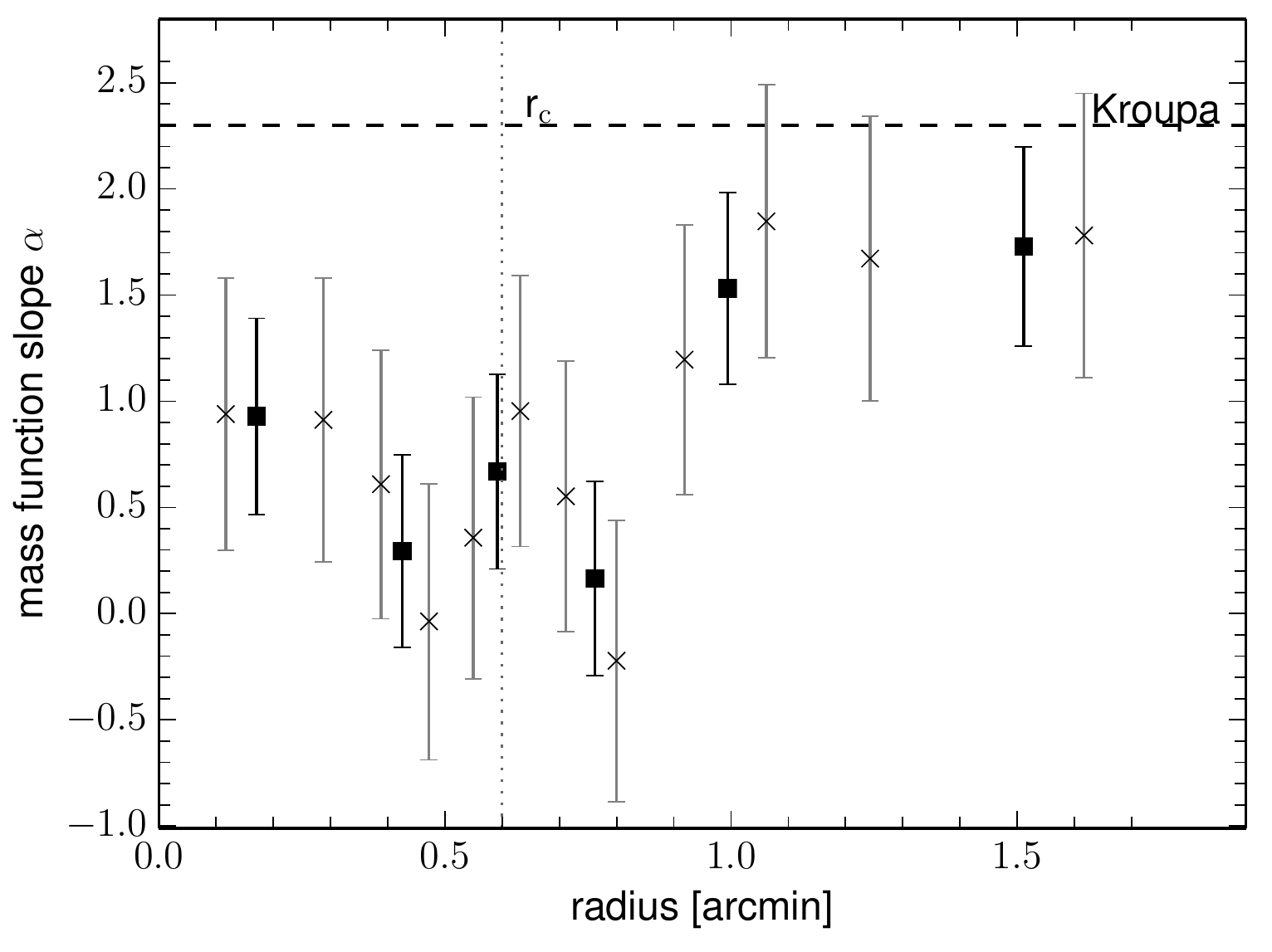}
\caption{The best-fitting mass function slope and its uncertainties in  six (black squares) and alternatively in 12 (gray crosses) statistically independent  radial bins containing equal numbers ( $354$, or $177$, respectively ) of observed stars. The radii of the data points correspond to the mean radius in the corresponding radial bin. The cluster's core radius $r_c=0.6$\,arcmin, or 12.4\,pc \citep{2011ApJ...726...47S} is indicated by the dotted line.  The increase  of  the mass function slope  $\alpha$ with increasing radius is obvious, although the slope stays below the \citet{2001MNRAS.322..231K} value of $\alpha=2.3$ in this range of masses (shown as dashed horizontal line) out to the maximum radius covered by the WFPC2 photometry.}
\label{fig:alphavsradius}
\end{figure}

In the combined radial bins, we sorted stars by mass into  9  mass bins with a width of $0.03\msun$ and fitted the resulting histograms with a power-law of the form $dN/dm\propto m^{-\alpha}$. It is common practice and well justified in the case of large number counts $n_{i}$, to treat their uncertainties as Gaussian with dispersions of $\sqrt{n_i}$. In this case, a maximum likelihood fit is obtained by minimizing $\chi^2$ or, equivalently, by maximizing the log-likelihood function 
$$\mathrm{ln}\,\mathcal{L} = k-\sum_{i=1}^{N}{\frac{\left(f_i - n_i\,c_i^{-1}\right)^2}{2\left(\sigma_i\,c_i^{-1}\right)^2}}~,$$ where $k$ is a constant, $f_i=\int_{m_i}^{m_{i+1}}{dN}$ is the model prediction in the $i$-th mass-bin (ranging from mass $m_i$ to $m_{i+1}$), $\sigma_i=\sqrt{n_i}$ is the uncertainty of the observed number of stars $n_i$ and $c_i$ is the completeness (photometric and geometric) in the given bin. It is also well-known that the Poissonian nature of number counts has to be taken into account in the case of small numbers \citep[e.g.][]{1999ApJ...518..380M}. Therefore, we chose to maximize the Poissonian log-likelihood function
$$\mathrm{ln}\,\mathcal{L} = -\sum_{i=1}^{N}{\mathrm{ln}\left(\frac{\left(\mathrm{f_i}\,c_i\right)^{n_i}\,\mathrm{exp}\left(f_i\,c_i\right)}{n_{i}!}\right)}~,$$ 
where the completeness correction has been multiplied into the model term instead of divided into the $n_i$.  Because of the large number of $\sim200000$ artificial stars, the statistical uncertainty of the completeness is $\ll1$\,per cent in most radial and mass bins and never exceeds $2.5$ per cent, and is therefore negligible compared to that of the observed star counts.

There are several methods to estimate confidence intervals from the likelihood function; one of them is the likelihood ratio method \citep[e.g.][]{Hudson1971}, which we used here: given the maximum likelihood, the $p$\,percent confidence interval for $k$ parameters of interest is defined as the contour in parameter space, where the likelihood function has dropped by a factor of $\mathrm{exp}\left(-Q_{\chi^2_k}\left(p\right)/2\right)$ with respect to its maximum. Here $Q_{\chi^2_k}\left(p\right)$ is the quantile function of the $\chi^2$ distribution for $k$ degrees of freedom; for one parameter of interest, the $68.3$\,percent (or $1\,\sigma$) confidence interval is given by a decrease in the likelihood by a factor of $\mathrm{exp}\left(-1/2\right)$. The confidence limits derived in this way in principle need not be symmetric around the maximum likelihood solution. Nevertheless, our calculated upper and lower uncertainties on $\alpha$ are almost identical, indicating that the likelihood function is symmetric around its maximum. This is also supported by the fact that the results and uncertainties obtained with this Poissonian maximum likelihood fitting did not significantly differ from those obtained for comparison via a common $\chi^2$-minimization, despite the low number counts.

The amount of radial sub-divisions is clearly a trade-off between signal-to-noise ratio and radial resolution and we present two realizations of the binning. Fig.~\ref{fig:masssegregation} shows the mass distributions of observed stars (red dotted lines), the distributions corrected for missing area (blue dashed lines) and the distributions additionally corrected for photometric completeness (black solid lines), as well as the best-fitting power-law mass function, in radial ranges containing each one fourth of the observed stars (i.e. combining blocks of 9 of the 36 original radial bins). The observed and completeness-corrected mass function in these ranges are also reported in Table~\ref{tab:massfuncs}. 
From Fig.~\ref{fig:masssegregation}, it can be seen that the mass function in each radial interval is well-reproduced by a single power-law with a slope shallower than a \citet[][]{2001MNRAS.322..231K} mass function. Moreover, the slope $\alpha$ in the outer-most radial range ($1\!<\!r\!\le\!1.9$\,arcmin, corresponding to $20.7\!<\!r\!\le\!39.2$\,pc) is significantly steeper than in the inner two bins. 

The increase of  $\alpha$ with increasing radius is more clearly seen in the finer radial samplings in  Fig.~\ref{fig:alphavsradius}, which shows the best-fitting mass function slope $\alpha$ as a function of radius for  statistically independent bins containing one sixth of  observed stars (black squares), and alternatively containing one 12th of observed stars (gray crosses). To quantify the radial dependence of $\alpha$, we fitted different possible models to the data shown as gray crosses in Fig.~\ref{fig:alphavsradius}: a constant, a linear relation, a bimodal relation (i.e. a piece-wise constant function with one discrete change), as well as a quadratic relation between radius and $\alpha$. We used standard $\chi^2$ minimization, i.e. assuming that the uncertainties on each $\alpha$ measurement are Gaussian, which is supported by the symmetry of the likelihood functions found around the best-fitting $\alpha$ values, as reported above. The functional form and best-fitting parameters for each model are listed in Table~\ref{tab:alphavsradius}. It is trivial that a model with more free parameters reproduces the fitted data more closely. On the other hand a, potentially over-fitted model with more free parameters also will do worse in predicting unknown data. Thus, in order to quantify the relative odds of the different models, we use Jackknife, or leave-one-out, cross-validation \citep[e.g.][]{Rust1995}: for each model $m$, each of the twelve data points is left out in turn, the maximum-likelihood parameters are determined from the remaining eleven data points. The likelihood $L_{m,\{i\}}$ of the left-out data point $i$ given this best-fitting relation is then evaluated. The cross-validated pseudo-likelihood for model $m$ is then $$L_{m}^{\rm CV}=\prod_{i=1}^{12}\,L_{m,\{i\}}.$$ Under the implicit assumptions of equal prior probabilities for all models and that the true model is contained in the set of examined models, the relative probability for model $M$ is then $$p_M=L_{M}^{\rm CV} \left( \sum_{m}^\mathrm{models} L_{m}^\mathrm{CV}\right)^{-1}.$$ From the relative probabilities, which are reported in the last column of Table~\ref{tab:alphavsradius}, the data favor a bimodal relation (with a probability of 88.3\,per cent). The data also allow for a linear rise of $\alpha$ with radius (10\,per cent), and suggest that a quadratic relation ($<0.1$\,per cent), and a constant relation, although less significantly ($1.7$\,per cent) can be excluded.

The apparent steepening (either discrete or continuous) of the mass function as a function of radius indicates  that mass segregation \emph{is} present in Pal\,14. Moreover, Fig.~\ref{fig:alphavsradius} shows that the mass function at the outermost radii covered by our photometry is almost compatible with a Kroupa IMF ($\alpha=2.3$ in this mass range); since it is likely that the relative abundance of low-mass stars increases further with increasing radius, this suggests that there is no need to invoke a deviation from a universal IMF in this cluster \citep[cf. ][]{2011MNRAS.411.1989Z}.

\begin{table*}
\caption[]{ Functional form of $\alpha(r)$ fitted to the data points shown as gray crosses in Fig.~\ref{fig:alphavsradius}. For each model, the number of free parameters, the functional form with best-fitting parameters and the relative probability of the model are given.}
\begin{tabular}{|l|r|c|r|r|}
\hline \hline
model & parameters & functional form & relative probability \\
\hline
constant & 1 & $\alpha(r) = 0.9\pm0.2 $ & 0.0171\\
linear & 2 & $ \alpha(r) = (0.2\pm0.4) + r\times (0.9\pm0.5) \mathrm{arcmin^{-1}}$ & 0.0996\\
quadratic & 3 & $ \alpha(r) = (0.8\pm0.7) + r\times (0.9\pm1.7) \mathrm{arcmin^{-1}} + r^2\times (1.1\pm1.0) \mathrm{arcmin^{-2}}$ & 0.0004\\
bimodal & 3 & $\alpha(r) = (0.5\pm0.2) (r\le0.86\pm0.05), 1.6\pm0.3 (r>0.86\pm0.05)$ & 0.8828\\
\hline \hline
\end{tabular}
\label{tab:alphavsradius}
\end{table*}

\subsection{Overall mass function}
 \label{sec:overallmassfunction}
 The overall mass function slope of stars within $1.9$\,arcmin (i.e. subsuming the star counts from all radial bins) is $\alpha=1.1\pm0.2$. This is somewhat shallower than the value of $\alpha=1.3\pm0.4$ derived by \citet{2009AJ....137.4586J} from the same data. Although consistent within the uncertainties, the difference is, at first glance, surprising, as we also adopted the best-fitting isochrone from \citet{2009AJ....137.4586J} in our analysis. The difference can be traced back to a difference in the completeness correction: we calculated the completeness based solely on artificial stars inside the color selection used for the observed stars. If we instead calculated the completeness from the entire artificial star catalog, we would obtain a larger completeness correction in the lowest mass bins, similar to the values given in table~3 of \citet{2009AJ....137.4586J}. However, as can be seen from the completeness contours in the CMD (Fig.~\ref{fig:CMD}), for faint stars bluer than the cluster's main sequence, detection is limited by the F814W filter. Therefore, the inclusion of artificial stars in that region of the CMD would result in an over-correction for completeness. 

 \subsection{Systematics in the measurement of the mass function}
\label{sec:massfunctionsystematics}
\subsubsection{Dependence on the adopted isochrone}
 To estimate the systematic effects of the adopted stellar population parameters on the inferred mass function, we measured the mass function using the same `measurement' isochrone as before from an artificial luminosity distribution that was generated based on the assumption that the cluster's actual stellar population parameters are different. For this experiment we chose an `input' isochrone from the \citep{2008ApJS..178...89D} data base that differed in age, [Fe/H], or [$\alpha$/Fe] from our adopted values. We then drew 2000 stars in the range of  $0.53-0.80\msun$  from a power-law mass function with slope $\alpha_\mathrm{in}$ and converted the masses to a luminosity distribution according to this input isochrone. 

The location of the main-sequence turn-off of the input isochrone in general differed from that of the measurement isochrone. However, any isochrone advocated as the best-fitting isochrone in an observational study will be one that reasonably well reproduces the cluster's main-sequence turn-off. Therefore, we shifted our measurement isochrone in magnitude and color such that its turn-off point coincided with the insertion isochrone. In observations, these shifts would be equivalent to differences in the inferred distance and reddening and possibly difference in the adopted magnitude zeropoints. 

Using this shifted measurement isochrone, we obtained the best-fitting slope of the mass function $\alpha_\mathrm{rec}$ from the artificial luminosity distribution in the same way as for the observations. We find that over- or underestimating the cluster's age, metallicity or [$\alpha$/Fe] by 1\,Gyr or 0.2\,dex respectively, results in systematic offsets between the inserted and measured mass function slope by $\Delta\alpha = \|\alpha_\mathrm{rec}-\alpha_\mathrm{in}\|=0.05-0.15$. The effect of solely the distance uncertainty, estimated by using a measurement isochrone with the same age, [Fe/H] and [$\alpha$/Fe] as the input isochrone, but shifting it by $\pm$0.04\,mag in distance modulus (corresponding to the distance uncertainty of 1.3\,kpc, cf. Section~\ref{sec:populationparameters}), yields  an  offset of $\Delta\alpha\leq0.05$. In the tested range of inserted mass function slopes, $\alpha_\mathrm{in}=0 - 3$, the offsets only weakly depend on the slope $\alpha_\mathrm{in}$ itself. Therefore, the amplitude of the change in the mass function slope at different radii of the cluster is largely independent of the adopted stellar population parameters and cluster distance, while the absolute value of the slope may be considered subject to a systematic uncertainty of $\la0.15$.

 \subsubsection{Unresolved binaries}
 \begin{figure}
 \includegraphics[width=84mm]{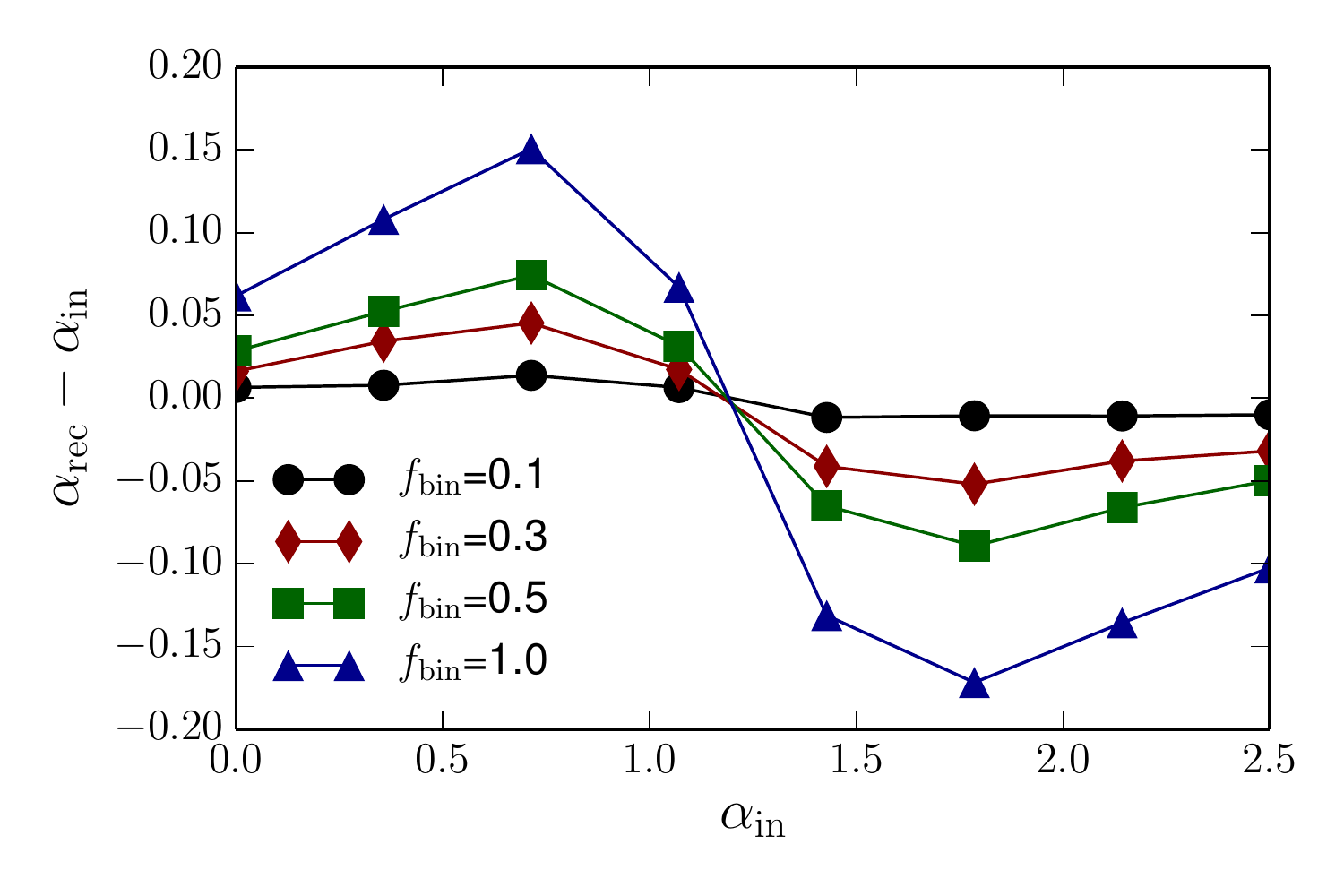}
 \caption{The  influence  of  unresolved binaries on  the  measured mass function slope.  The difference of recovered $\alpha_\mathrm{rec}$  and  inserted $\alpha_\mathrm{in}$ is shown as a function of $\alpha_\mathrm{in}$ based on simulations  (see text)  including a varying fraction  of  binaries,  $f_\mathrm{bin}$ ranging from 0.1  to  1  and  represented by different symbols.  For realistic binary fractions  of  $f_\mathrm{bin}\la0.5$,  the  impact  of  randomly paired binaries is relatively small, with a maximum systematic uncertainty in $\alpha$ of $\pm0.07$.}
 \label{fig:effectofbinaries}
 \end{figure}

 In measuring the mass function, we do not take into account the effect of unresolved binaries that are likely contaminating our main-sequence star sample. As mentioned in  the  introduction, dynamical constraints allow for a binary fraction of $f_\mathrm{bin}\sim0.3$ \citep{2012ApJ...744..196S} in Pal\,14. From the specific frequency of blue stragglers and  the  total magnitude of  the  cluster, \citet{2011ApJ...737L...3B} estimated a similar binary fraction of $f_\mathrm{bin}\sim0.3-0.4$.

To estimate the impact of the presence of binaries that are counted as single stars in our measurement, we drew stellar masses from a power-law distribution with varying slope $\alpha_\mathrm{in}$ in the mass-range $0.08-1\msun$. We randomly paired a fraction of $f_\mathrm{bin}/(1+f_\mathrm{bin})$ of these masses, and for each pair calculated the sum of the fluxes of both components from our adopted isochrone and converted the summed flux back to a mass. This results in a sample of stellar masses, in which a fraction of $f_\mathrm{bin}$ represent the mass that would be inferred for an unresolved binary. We measured the best-fitting mass function slope from the sub-set generated stars/binaries with (apparent) masses in the range $0.53-0.8\,\msun$. Fig~\ref{fig:effectofbinaries} shows the difference of the recovered mass function slope $\alpha_\mathrm{rec}$ and the input mass function slope $\alpha_\mathrm{in}$ as a function $\alpha_\mathrm{in}$ for binary fractions $f_\mathrm{bin}=0.1, 0.3, 0.5$, and as an (unrealistic) limiting case, $f_\mathrm{bin}=1$. Realistic binary fractions ($f_\mathrm{bin}\la0.5$) result in a relatively small systematic error in  the  mass function slope, $|\alpha_\mathrm{rec}-\alpha_\mathrm{in}|\la0.07$. It is interesting to note that in the presence of unresolved binaries, mass functions with intrinsic slopes of $\alpha_\mathrm{in}\sim0.7$ tend to appear steeper, while mass functions with intrinsic slopes of $\alpha_\mathrm{in}\sim1.7$ tend to appear flatter. Thus, an observed change in the mass function slope from $\sim0.5$ to $\sim1.6$ like we see it in Pal\,14, might in reality even be somewhat larger than suggested by a measurement that ignores possible binarity. In the case of our data, however, the effect would be much smaller than the statistical uncertainty.

\subsubsection{Photometric errors}
\label{subsubsec:photerrors}
Photometric errors, i.e.  statistical uncertainties on measured magnitudes, as well as systematics caused by blending or imperfect subtraction of neighboring stars, directly translate into an error in the inferred mass of a star. To estimate the impact of these errors, we followed a similar procedure: we drew stars from a mass function with $\alpha_\mathrm{in}=0 - 3$ and converted the masses to magnitudes using the adopted isochrone. For each generated star, we then randomly chose one of the 100 artificial stars (described in Section~\ref{sec:obs}) nearest in terms of inserted F555W magnitude to our generated magnitude. If that artificial star was recovered, we added to our generated magnitude the offset between recovered and inserted magnitude of the artificial star (see Fig.~\ref{fig:photerrors}). If the artificial was not recovered, we rejected the generated magnitude in order to simulate the same degree of incompleteness as in our observations. We converted the resulting sample of magnitudes back to masses and measured the mass function slope $\alpha_\mathrm{rec}$ in the same mass bins and using the same completeness correction as for the observations. We find that photometric errors make the mass function appear systematically more steep, although the offset $\alpha_\mathrm{rec}$-$\alpha_\mathrm{in}$ is marginal, decreasing slightly from 0.08 for $\alpha_\mathrm{in}=0$ to 0.03 for $\alpha_\mathrm{in}=3$. Compared to the statistical uncertainties of our measured mass function slopes, this effect is negligible. 

\section{Discussion}
\label{sec:disc}
We have shown that the mass function in Pal\,14 steepens with increasing distance from the center, which suggests  that Pal\,14 is mass-segregated. 

\subsection{Mass segregation vs. the large extent of the cluster}
\citet{2011ApJ...726...47S} measured a projected half-light radius for Pal\,14 of $r_{hp}=46\pm3$\,pc, making Pal\,14 the most extended Galactic GC and corresponding to a present-day half-mass relaxation time for Pal\,14 of $\sim20$\,Gyr. As noted before, it is therefore not expected that two-body relaxation processes have established mass segregation in Pal\,14, unless the cluster had a shorter relaxation time in the past. Since the half-mass relaxation time-scales with r$_{hp}^{3/2}$ \citep{1971ApJ...164..399S}, a half-mass relaxation time of a few Gyr would require the cluster to have been more compact by a factor of $\sim\!2$ for most of its lifetime, or even more compact for a shorter duration. 

Since the stellar distribution of Pal\,14 extends out to more than twice of its Jacobi radius of $r_J=170\pm10$\,pc \citep[][assuming a circular cluster orbit about the Galaxy]{2011ApJ...726...47S}, it is plausible that the cluster was  indeed  more compact in the past, or otherwise it is puzzling how such a diffuse cluster could have survived until the present day in the Galactic tidal field. Alternatively, it may have evolved in a less hostile environment of a satellite galaxy and may have been only recently accreted \citep{2011ApJ...726...47S}. \citet{2012A&A...537A..83C} concluded that the element abundance ratios of red giants in Pal\,14 are consistent with, but do not require an accretion origin of the GC from a dwarf spheroidal galaxy.

In the latter case, i.e. the cluster was similarly extended, and therefore had a similar relaxation time-scale in the past, the observed mass segregation may be primordial. Several mechanisms that can lead to a primordial, or early, mass segregation in clusters have been proposed. \citet{2007ApJ...655L..45M} propose a scenario, in which young star clusters assemble from less massive clumps that have significantly shorter relaxation time-scales, and which therefore can be already mass-segregated at the time of the cluster assembly. The resulting merged cluster will inherit the mass segregation from its constituents. \citet{2011IAUS..271..389O} found that early dynamical mass segregation also occurs in clusters without substructure, if they are initially sub-virial and therefore undergo a cold collapse. Another possibility is that mass segregation is created in the star-formation process itself due a more efficient accretion of gas through protostars that reside in the cluster's center, where the density of gas is higher  \citep[competitive accretion, see][]{2001MNRAS.323..785B,2001MNRAS.324..573B}. 

Primordial mass segregation is observed in several young Galactic \citep[e.g.][]{1988MNRAS.234..831S,1997AJ....113.1733H,2011MNRAS.413.2345H}
and Magellanic Cloud star clusters \citep[e.g.][]{1998AJ....115..592F,2002ApJ...579..275S}. Besides Pal\,14, also the globular clusters Pal\,5 \citep{2004AJ....128.2274K}, Pal\,4 \citep{2012MNRAS.423.2917F} and potentially the rich cluster Lindsay\,38 in the Small Magellanic Cloud \citep{2011AJ....142...36G} show mass segregation despite having present-day half-mass relaxation times larger than their ages. \citet{2008ApJ...685..247B} suggested that primordial mass segregation together with a depletion of low-mass stars are required to explain the 
shallow present-day mass functions of at least some Galactic GCs. As mentioned in the introduction, this was also found by \citet{2011MNRAS.411.1989Z} for Pal\,14.

For Pal\,4, the observational data resemble our present results for Pal\,14. Pal\,4's half-mass relaxation time \citep[14~Gyr;][]{2012MNRAS.423.2917F}, albeit shorter than that of Pal\,14, exceeds its age, and yet the cluster shows a flattened mass function, as well as a substantial degree of mass segregation in the mass range of $0.55-0.8\msun$. \citet{2014MNRAS.440.3172Z} presented direct $N$-body simulations of Pal\,4 to specifically address the question whether or not primordial mass segregation is necessary to reproduce the observations. They studied the evolution over 11\,Gyr of different models of GCs on circular orbits, starting with either a Kroupa or an initially flattened IMF, and either with or without primordial mass segregation. They found that the current state of Pal\,4 (i.e. its mass function, mass segregation, half light radius and mass) can be reproduced \emph{either} by assuming primordial mass segregation, \emph{or} by assuming an IMF already depleted in low-mass stars, compared to the Kroupa IMF. 

Similarly, using collisional $N$-body simulations to study Pal\,14, \citet{2011MNRAS.411.1989Z} have shown that two-body relaxation alone is insufficient in order to deplete the mass function of low-mass stars and hence to establish the observed mass segregation, if the cluster was born with an initial half-mass radius similar to its present-day value and with a Kroupa IMF. On the other hand, they found that a cluster born initially significantly more compact will not evolve to the present extended shape of Pal\,14. This is consistent with Pal\,14 being (by a factor of $\sim\!4$) more extended than predicted by the mass-radius relation that was found by \citet{2010MNRAS.408L..16G}. This relation can reproduce the size of the majority of Galactic GCs based on the assumption of initially compact clusters whose expansion during 10\,Gyr of evolution is driven by mass loss due to stellar evolution, as well as a central energy source such as hard binaries.  Both theoretical results suggest that an initially compact cluster, in which mass segregation would have developed dynamically, cannot have expanded to the observed present size of Pal\,14 due to \emph{internal} processes alone.

On the other hand the tidal tails around the cluster show that it is subject to \emph{external} processes even at its current large Galactocentric distance of $66$\,kpc. \citet{2010MNRAS.401.1832B} used the ratio of the de-projected half-mass radius $r_h$ to $r_J$ to quantify how tidally filling a GC is and found two distinct populations among the Galactic GCs, a tidally under-filling group with $r_h/r_J<0.05$ and a tidally filling group with $0.1<r_h/r_J<0.3$. Using $r_h=4/3\times r_{hp}$ and neglecting mass segregation by equating the half-light to the half-mass radius, yields a ratio $r_h/r_J=0.35$ for Pal\,14, making it one of the most tidally filling Galactic GCs, second only to Pal\,5. The extended structure of Pal\,5 was explained by \citet{2004AJ....127.2753D} as the result of its expansion due to heating induced by a tidal shock during a recent passage through the Galactic disk. Disk shocks in general dominate the evolution of diffuse clusters in the inner Galactic potential \citep{1999ApJ...522..935G} and a significant fraction ($\ga17$ per cent) of the inner halo field stars may have originated in GCs \citep{2011A&A...534A.136M}. Pal\,5 is located at a Galactocentric distance of $18.5$\,kpc and is believed to be near its apogalacticon \citep{2003AJ....126.2385O}. Pal\,14 has a current Galactocentric distance 3.5 times as large, but its orbit is unknown. 

In order for disk shocks to become important, a cluster has to cross the Galactic disk within a Galactocentric radius of about 8 kpc, since disk shocks will be considerably weaker for a cluster crossing the outer disk where the stellar density is much lower \citep{1997MNRAS.289..898V}. In addition, disk crossings will occur less frequently for a cluster on a nearly circular orbit at the Galactocentric distance of Pal\,14. So Pal\,14 would have to be on an eccentric orbit. In fact, assuming that its current distance of 66 kpc is close to its apogalactic radius, $R_{apo}$, Pal\,14 would have to be on a highly eccentric orbit with $\epsilon = (R_{apo}-R_{peri})/(R_{apo}+R_{peri}) = (66-8)/(66+8) \approx 0.8$ in order to have a perigalactic radius, $R_{peri}$, below 8 kpc. Such high eccentricities would imply severe tidal shocking during pericentre passages, being much more hazardous to the cluster than the disk shocks. Such tidal shocks could in fact temporarily unbind large parts of a cluster like Pal\,14, since its Jacobi radius could be significantly smaller than the size of the cluster during a pericentre passage. While moving towards apogalacticon the Jacobi radius of the cluster would grow again and re-capture large parts of the temporarily unbound material \citep{2010MNRAS.401..105K}.

It is therefore possible that the half-light radius of Pal\,14 is significantly overestimated due to unbound stars. In $N$-body simulations of GCs on eccentric orbits in external tidal fields, \citet{2010MNRAS.407.2241K} found that beyond $0.5\,r_J$, the surface density profiles of the clusters can be entirely dominated by unbound stars. Since Pal\,14 may not be on a circular orbit and may not currently be at its perigalacticon, the mean Jacobi radius to which the cluster adjusts \citep{2010MNRAS.407.2241K} may be smaller than the present-day value of $r_J=170\pm10$\,pc. Therefore the distribution of stars attributed to Pal\,14 could already be dominated by unbound stars at radii smaller than $\la\!85$\,pc, and the half-light radius of the bound body of Pal\,14 may be significantly smaller than the estimated half-light radius of 46\,pc. In this case, the mystery how the diffuse cluster has survived to the present day has a simple solution: the cluster has \emph{not} survived, and what we observe is an already largely disintegrated stellar system.   

\subsection{Mass segregation vs. the non-concentrated blue straggler population}
 As noted in the introduction, \citet{2011ApJ...737L...3B}, based on a geometrically and photometrically complete catalog, found that the population of 24 BS in Pal\,14 is not centrally concentrated compared to 24 HB and 191 selected RGB stars. 
 Other Galactic globular clusters, for which a non-segregated population of BS stars (compared to HB or RGB stars) has been found are NGC\,5139 \citep[$\omega$Cen;][]{2006ApJ...638..433F}, NGC\,2419 \citep{2008ApJ...681..311D}, Terzan\,8 \citep{2012MNRAS.421..960S} and Arp\,2 \citep{2012MNRAS.421..960S,2011MNRAS.412.1361C}. It is interesting to note that an extragalactic origin has been suggested for all of these GCs: Terzan\,8 and Arp\,2 are likely associated with the Sgr stream  \citep{1995AJ....109.2533D,2010ApJ...718.1128L}. $\omega$Cen is commonly seen as the remnant nucleus of a former Milky Way satellite galaxy \citep[e.g.][]{1999Natur.402...55L,2000LIACo..35..619M,2000A&A...362..895H,2003MNRAS.346L..11B} and a similar scenario is being discussed for NGC\,2419, although the case seems less clear for this GC \citep[e.g.][]{2004MNRAS.354..713V,2007ApJ...667L..61R,2010ApJ...725..288C,2011ApJ...729...69B,2011MNRAS.414.3381D}.

For $\omega$Cen, \citet{2007MNRAS.381.1575S} found no difference in the luminosity function of stars in fields at 12 and 20\,arcmin from the cluster center. There are also no indications for a significantly varying dynamical mass-to-light ratio out to $\sim3$ half-light radii \citep{2006A&A...445..513V,2012A&A...538A..19J} and also the velocity dispersion of MS stars as a function of their mass indicates that energy equipartition has not been established in the cluster's center \citep{2010ApJ...710.1032A}. 
Also for NGC\,2419 there is no evidence for a radially changing luminosity function \citep{2012MNRAS.423..844B}. Therefore, $\omega$Cen and NGC\,2419 can be regarded as not being mass-segregated. For Terzan\,8 no further data beyond the comparison of BS and HB and RGB stars by \citet{2012MNRAS.421..960S} are available. In Arp\,2, which has the lowest present-day half-mass relaxation time among these GCs (8\,Gyr, \citealt{2012MNRAS.421..960S}), the situation is similar to  the situation in  Pal\,14:  \citet{2011MNRAS.412.1361C} find that  the radial distribution of BS stars is not  statistically  different from that of HB and RGB stars  (although the histogram shown in their fig. 6 may convey the impression of a higher concentration of BS in the innermost radial bin), but MS stars have a more extended radial distribution, indicating that Arp\,2 is mass-segregated.

The apparent conflict in clusters like Pal\,14 and  Arp\,2 between a non-concentrated population  of BS and  a MS population that shows mass segregation may be due to the different time-scales traced by the sinking-in of BS and the depletion of MS stars in the center. The time-scale for mass segregation is shortest for the highest mass stars, turning into the dynamical friction time-scale in the case of masses much larger than the average stellar mass in the cluster. Therefore at some early point in the evolution of a cluster, the highest mass stars, or rather their stellar remnants, will already have sunken to the center, while the intermediate-mass BS still remain largely unaffected by dynamical evolution. The orbital energy lost by the high-mass remnants will deplete the center of, preferentially, low-mass stars, such as those on the main sequence.  Indeed, simulations of Pal\,4 \citep{2014MNRAS.440.3172Z} show that on a time-scale of 11\,Gyr, which is of the same order of magnitude as Pal\,4's half-mass relaxation time, two-body relaxation processes are sufficient to establish an observable degree of mass segregation among main sequence stars in the mass range $0.55-0.8\,\msun$ -- even in GCs, which have low masses and large half-light radii at the present day. In the framework of \citet{2012Natur.492..393F}, who classify Galactic GCs into three stages of dynamical evolution based on the radial distribution of BS, Pal\,14, together with $\omega$Cen and NGC\,2419 comprises the dynamically unevolved `family 1' of GCs. Since in contrast to $\omega$Cen and NGC\,2419, Pal\,14 is already depleted of low-mass stars in the center, in this picture, Pal\,14 might be on the verge of transitioning towards the dynamically intermediate `family 2'.

Finally, we note that in the competitive accretion scenario that was mentioned above in the context of primordial mass segregation, it  is also  conceivable that stars are generally segregated by mass, but that the distribution of binaries (such as the BS progenitors) is not necessarily more centrally concentrated, although to our knowledge this has not been studied theoretically yet.

\section{Summary}
\label{sec:sum}
 Using archival \textit{HST} data we showed  that the main-sequence stars of Pal\,14 clearly indicate mass segregation in the sense that the cluster is significantly depleted in low-mass stars in its core with respect to larger radii. 

The radial dependence of the stellar mass function in Pal\,14 was measured using deep archival $V$ and $I$-band imaging obtained with the \textit{HST}/WFPC2. We find that in the stellar mass range  $0.53\leq m\le0.80\msun$  the stellar mass function is described by a shallow power-law slope of  $\alpha=0.5\pm0.2$  in the cluster's core, while at larger radii the present-day mass function with a slope of  $\alpha=1.6\pm0.3$ is still below  a Kroupa IMF in this mass range ($\alpha$=2.3).

In order for mass segregation in Pal\,14 and its overall mass function being depleted of low-mass stars to be the result of two-body relaxation processes, Pal\,14 must have been significantly more compact in the past, since its present-day half-mass two-body relaxation time is of the order of 20~Gyr. The expansion of more than a factor of 2 in size may have been caused by severe tidal shocks during pericentre passages on a very eccentric cluster orbit about the Milky Way, as it has been similarly suggested for Palomar 5 with disk shocks \citep{ 2004AJ....127.2753D}.

Alternatively, the cluster might have been given this extended shape by the gas expulsion process at birth which might also be the origin of the depleted mass function \citep{2008MNRAS.386.2047M}. For such an extended cluster to have survived to the present day, this birth process is likely to have happened in a very remote environment, far away from the destructive tidal forces of the Milky Way. 

Hence, Pal\,14 could either be an initially compact native Milky Way cluster on a very eccentric orbit, or be a recently accreted globular cluster which formed with primordial mass segregation far away from the Galaxy.

In any case, due to the mass segregation the tidal tails of Pal\,14 should show a mass function enriched in low-mass stars, so that mapping their full extent will require deep imaging reaching well below the cluster's main-sequence turn-off point.

\section*{Acknowledgments}
This work was partially supported by Sonderforschungsbereich 881, ``The Milky
Way System'' (subprojects A2 and A3) of the German Research Foundation (DFG) at
the University of Heidelberg.

M.J.F.  and A.H.W.K. gratefully acknowledge support from the DFG via Emmy Noether Grant Ko 4161/1 and project KR 1635/28-1, respectively. AHWK would like to acknowledge support through DFG Research Fellowship KU 3109/1-1 and from NASA through Hubble Fellowship grant HST-HF-51323.01-A awarded by the Space Telescope Science Institute, which is operated by the Association of Universities for Research in Astronomy, Inc., for NASA, under contract NAS 5-26555.

Based on observations made with the NASA/ESA Hubble Space Telescope, obtained
from the Multimission Archive at the Space Telescope Science Institute (MAST).
STScI is operated by the Association of Universities for Research in Astronomy,
Inc., under NASA contract NAS5-26555.

\bsp

\label{lastpage}

\end{document}